# A framework for fast quantum mechanical algorithms

Lov K. Grover
3C-404A, Bell Labs
600 Mountain Avenue
Murray Hill NJ 07974
*lkgrover@bell-labs.com*

**Summary**

A framework is presented for the design and analysis of quantum mechanical algorithms, the $O(\sqrt{N})$ step quantum search algorithm is an immediate consequence of this framework. It leads to several other search-type applications - an example is presented where the Walsh-Hadamard (W-H) transform of the quantum search algorithm is replaced by another transform tailored to the parameters of the problem. Also, it leads to quantum mechanical algorithms for problems not immediately connected with search - two such algorithms are presented for calculating the mean and median of statistical distributions. In order to classically estimate either the mean or median of a given distribution to a precision $\varepsilon$, needs $\Omega(\varepsilon^{-2})$ steps. The best known quantum mechanical algorithm for estimating the median takes $O(\varepsilon^{-1})$ steps, and that for estimating the mean takes $O(\varepsilon^{-1.5})$ steps. This paper presents $O(\varepsilon^{-1})$ step algorithms for both problems (all bounds are upto polylogarithmic factors). Both algorithms are considerably simpler than known algorithms.

## 1. Introduction

**1.0 Background** Quantum computers were first considered in the 70's and early 80's. In 1980, it was shown that anything that could be computed by a classical computer could, in principle, be computed by a quantum computer [Benioff]. But could a quantum computer do anything that a classical computer could not? Feynman [Feynman] discovered an example of a quantum mechanical system which seemed very difficult to simulate on a classical computer, but which a quantum computer could easily do in polynomial time.

Through the late 80's and early 90's, the description of quantum computers was formalized and a number of contrived situations and oracle-based examples were discovered that a quantum computer could solve more efficiently than a classical computer [DJ][Yao] [Simon][BV].

The question still remained as to whether a quantum computer could solve an actual problem of interest to computer scientists more efficiently than a classical computer. Building a quantum computer is a challenging task and before attempting to build one, the designers would need to be convinced of its ultimate usefulness. In 1994, Peter Shor presented an efficient quantum mechanical algorithm for factorization - this was a problem of great interest to computer scientists and it aroused a lot of excitement [Factor]. It was expected that this would be quickly followed by other quantum mechanical algorithms; however, the next significant result had to wait till 1996 when an efficient algorithm for exhaustive search was discovered [Search1]. Since then fast quantum algorithms for other important computer science problems have been discovered [Collision][Mean][Median][Struct], but there is still no general technique for obtaining efficient quantum algorithms.

**1.1 This paper** This paper gives a general technique for deriving a class of fast quantum mechanical algorithms. The idea is to first consider a single quantum mechanical operation, or a combination of such operations, due to which there is a certain probability for the system to reach a certain target state *t*. It is shown that the probability of reaching the target state can be made to grow quadratically with the number of iterations by iterating the quantum mechanical operations in the prescribed way.

The most obvious application of this technique is exhaustive search. In the exhaustive search problem, a function $f(x), x = 1, 2 \ldots N$, is given which is known to be non-zero at a single (unknown) value of $x$, say $x_0$ - the goal is to find $x_0$. If there was no other information about $f(x)$ and one were using a classical computer, it is easy to see that on the average it would take $\frac{N}{2}$ func-tion evaluations to solve this problem successfully. However, quantum mechanical systems can explore multiple states simultaneously and there is no clear lower bound on how fast this could be done. [BBBV] showed by using subtle arguments about unitary transforms that it could not be done in fewer than $\Omega(\sqrt{N})$ steps - subsequently [Search1] found an algorithm that took precisely $O(\sqrt{N})$ steps. [Search1] was derived using the W-H transform as the state transition matrix and it appeared to be a consequence of the special properties of this transform, this paper shows that similar results are obtained by substituting *any* unitary transformation in place of the W-H transform. This flexibility leads to other search-type algorithms.

It also leads to applications that, at first sight, are not connected to search - two such applications to statistics are presented. In these a unitary transformation is designed due to which the amplitude in a particular state comes out to be proportional to the statistic we want to estimate. Then by repeating this transformation in the prescribed manner, the probability in the desired state is increased to a detectable level. Finally a measurement is made to determine if the system is indeed in the desired state. By repeating the entire operation sequence a few times and counting the number of observations, one can estimate the probability of occurrence of the desired state and hence the original statistic.

## 2. Quantum operations

In a quantum computer, the logic circuitry and time steps are classical, only the memory *bits* that hold the variables are in quantum superpositions - these are called *qubits*. Quantum mechanical operations that can be carried out in a controlled way are unitary operations that act on a small number of qubits in each step. A good starting point to think of quantum mechanical algorithms is probabilistic algorithms [BV] (e.g. simulated annealing). In these algorithms, instead of having the system in a specified state, it is in a distribution over various states with a certain probability of being in each state. At each step, there is a certain probability of making a transition from one state to another. The evolution of the system is obtained by premultiplying this probability vector (that describes the distribution of probabilities over various states) by a state transition matrix. Knowing the initial distribution and the state transition matrix, it is possible in principle to calculate the distribution at any instant in time.

Just like classical probabilistic algorithms, quantum mechanical algorithms work with a probability distribution over various states. However, unlike classical systems, the probability vector does not completely describe the system. In order to completely describe the system we need the *amplitude* in each state which is a complex number. The evolution of the system is obtained by premultiplying this amplitude vector (that describes the distribution of amplitudes over various states) by a transition matrix, the entries of which are complex in general. The probabilities in any state are given by the square of the absolute values of the amplitude in that state. It can be shown that in order to conserve probabilities, the state transition matrix has to be unitary [BV]. The machinery of quantum mechanical algorithms is illustrated by discussing two elementary unitary operations - these are the W-H transformation operation and the selective rotation of the amplitudes of certain states.

A basic operation in quantum computing is the operation $M$ performed on a single qubit - this is represented by the following unitary matrix: $M \equiv \frac{1}{\sqrt{2}}\begin{bmatrix} 1 & 1 \\ 1 & -1 \end{bmatrix}$ - the state 0 is transformed into the superposition: $\left(\frac{1}{\sqrt{2}}, \frac{1}{\sqrt{2}}\right)$. Similarly state 1 is transformed into the superposition $\left(\frac{1}{\sqrt{2}}, -\frac{1}{\sqrt{2}}\right)$. A system consisting of $n$ qubits has $N = 2^n$ basis states (in this paper whenever "a state of the system" is mentioned, it will mean one of the $N$ basis states in which each qubit is either a $0$ or a $1$). We can perform the transformation $M$ on each qubit independently in sequence thus changing the state of the system. The state transition matrix representing this operation will be of dimension $2^n \times 2^n$. Consider a case when the starting state is one of the $2^n$ basis states, i.e. a state described by a general string of $n$ binary digits composed of some 0s and some 1s. The result of performing the transformation $M$ on each qubit will be a superposition of states consisting of all possible $n$ bit binary strings with amplitude of each state being $\pm 2^{-\frac{n}{2}}$. This transformation is referred to as the W-H transformation [DJ] and denoted by $W$. This operation (or a closely related operation called the Fourier Transformation [Factor]) is one of the things that makes quantum mechanical algorithms more powerful than classical algorithms and forms the basis for most significant quantum mechanical algorithms.

The other transformation that we need is the selective rotation of the phase of the amplitude in certain states. The transformation matrix describing this for a 2 state system is of the form: $\begin{bmatrix} e^{i\phi_1} & 0 \\ 0 & e^{i\phi_2} \end{bmatrix}$ where $i = \sqrt{-1}$ and $\phi_1, \phi_2$ are arbitrary real numbers. Unlike the W-H transformation, the probability in each state stays the same.

The selective inversion of the phase of the amplitude in certain states is a particular case of selective rotation that we will need in all of the algorithms in this paper. Based on [BBHT], the following is a realization. Assume that there is a binary function $f(x)$ that is either $0$ or $1$. Given a superposition over states $x$, it is possible to design a quantum circuit that will selectively invert the amplitudes in all states where $f(x) = 1$. This is achieved by appending an ancilla bit, $b$ and considering the quantum circuit, as shown in figure below, that transforms a state $|x, b\rangle$ into $|x, f(x) XOR\, b\rangle$ (such a circuit exists since, as proved in [Revers], it is possible to design a quantum mechanical circuit to evaluate any function $f(x)$ that can be evaluated classically.) If the bit $b$ is initially placed in a superposition $\frac{1}{\sqrt{2}}(|0\rangle - |1\rangle)$, this circuit will invert the amplitudes precisely in the states for which $f(x) = 1$, while leaving amplitudes in other states unchanged.

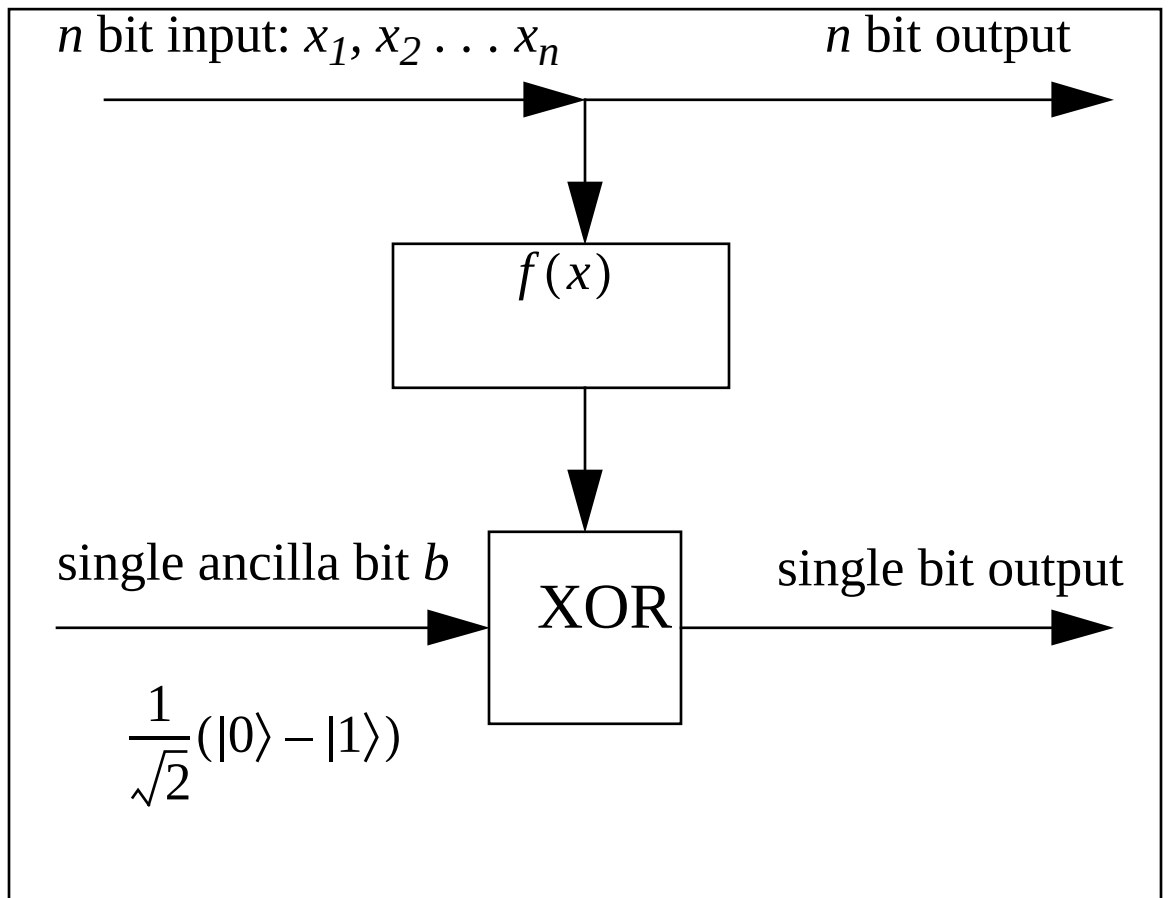

**Figure: The above quantum mechanical circuit inverts the amplitudes of precisely those states for which the function *f(x)* is 1.**

**3. Amplitude amplification**

Let each point of the domain of $f(x)$ be mapped to a state - let *t* be the target state, i.e. the function $f(x)$ is non-zero at the point corresponding to state *t*. The object is to get the system into the *t*-state. Assume that we have at our disposal a unitary transformation $U$ and we start with the system in the *s*-state. If we apply $U$ to *s*, the amplitude of reaching $t$ is $U_{ts}$, and if we were to observe the system at this point, the probability of getting the right state would be $|U_{ts}|^2$. It would therefore take $\Omega\left(\frac{1}{|U_{ts}|^2}\right)$ repetitions of this experiment before a single success. This section shows how it is possible to reach state *t* in only $O\left(\frac{1}{|U_{ts}|}\right)$ steps. This leads to a sizable improvement in the number of steps if $|U_{ts}| \ll 1$.

Denote the unitary operation that inverts the amplitude in a single state $x$ by $I_x$. In matrix notation this is the diagonal matrix with all diagonal terms equal to $1$, except the $xx$ term which is $-1$.

$v_x$ denotes the column vector which has all terms zero, except for the $x^{th}$ term which is unity.

Consider the following unitary operator[1]: $Q \equiv -I_s U^{-1} I_t U$. Note that since $U$ is unitary, $U^{-1}$ is equal to its adjoint, i.e. its conjugate transpose. We first show that $Q$ preserves the two dimensional vector space spanned by the two vectors: $v_s$ and $(U^{-1} v_t)$ (note that in the situation of interest, when $U_{ts}$ is small, the two vectors are almost orthogonal).

First consider $Qv_s$. By the definition of $Q$, this is: $-I_s U^{-1} I_t U v_s$. Note that $v_x v_x^T$ is an $N \times N$ square matrix all of whose terms are zero, except the $xx$ term

1. According to the notation of this paper (and most quantum mechanics texts), the operation $AB$ denotes that the sequence of operations is $B$ *and then* $A$. $Q \equiv -I_s U^{-1} I_t U$ implies the following operation sequence: First $U$, then $I_t$, then $U^{-1}$ and then $-I_s$.

which is 1 . Therefore $I_t \equiv I - 2v_t v_t^T$ & $I_s \equiv I - 2v_s v_s^T$ :

(1) $$Qv_s = -(I - 2v_s v_s^T)U^{-1}(I - 2v_t v_t^T)Uv_s$$
$$= -(I - 2v_s v_s^T)U^{-1}Uv_s + 2(I - 2v_s v_s^T)U^{-1}(v_t v_t^T)Uv_s$$

Using the fact that $v_s^T v_s \equiv 1$ , it follows that:

(2) $$Qv_s = v_s + 2(I - 2v_s v_s^T)U^{-1}(v_t v_t^T)Uv_s$$

Simplifying the second term of (2) by the following identities: $v_t^T U v_s \equiv U_{ts}$ and $v_s^T U^{-1} v_t \equiv U_{ts}^*$ (the second follows from the fact that $U$ is unitary and so its inverse is equal to its adjoint.)

(3) $$Qv_s = v_s(1 - 4|U_{ts}|^2) + 2U_{st}(U^{-1}v_t)$$

Next consider the action of the operator $Q$ on the vector $U^{-1}v_t$. Using the definition of $Q$ (i.e. $Q \equiv -I_s U^{-1} I_t U$) and carrying out the algebra as in the computation of $Qv_s$ above, this yields:

(4) $$Q(U^{-1}v_t) \equiv -I_s U^{-1} I_t U(U^{-1}v_t)$$
$$= -I_s U^{-1} I_t v_t = I_s U^{-1} v_t$$

Writing $I_s$ as $I - 2v_s v_s^T$ and as in (3), $v_s^T U^{-1} v_t \equiv U_{ts}^*$ :

(5) $$Q(U^{-1}v_t) = (U^{-1}v_t) - 2v_s v_s^T(U^{-1}v_t)$$
$$= (U^{-1}v_t) - 2U_{ts}^* v_s$$

It follows that the operator $Q$ transforms any superposition of the vectors $v_s$ & $(U^{-1}v_t)$ into another superposition of the two vectors, thus preserving the two dimensional vector space spanned by the two vectors $v_s$ & $(U^{-1}v_t)$. As indicated in the figure below, (3) & (5) may be written as:

(6) $$Q\begin{bmatrix} v_s \\ U^{-1}v_t \end{bmatrix} = \begin{bmatrix} (1 - 4|U_{ts}|^2) & 2U_{ts} \\ -2U_{ts}^* & 1 \end{bmatrix}\begin{bmatrix} v_s \\ U^{-1}v_t \end{bmatrix}$$

It follows as in [BBHT], that if we start with $v_s$, then after $\eta$ repetitions of $Q$ we get the superposition $a_s v_s + a_t U^{-1} v_t$ where $a_s = \cos(2\eta|U_{ts}|)$ and $|a_t| = |\sin(2\eta|U_{ts}|)|$. If $\eta = \frac{\pi}{4|U_{ts}|}$, then we get the superposition $U^{-1}v_t$, from this with a single application of $U$ we can get $v_t$. Therefore in $O\left(\frac{1}{|U_{ts}|}\right)$ steps, we can start with the *s*-state and reach the target state *t* with certainty.

The above derivation easily extends to the case when the amplitudes in states, *s* and *t*, instead of being inverted by $I_s$ and $I_t$, are rotated by an arbitrary phase. However, the number of operations required to reach *t* will be greater. Given a choice, it would be clearly better to use the inversion rather than a different phase rotation. Also the analysis can be extended to include the case where $I_t$ is replaced by $V^{-1}I_tV$, $V$ is an arbitrary unitary matrix. The analysis is the same as before but instead of the operation $U$, we will now have the operation $VU$.

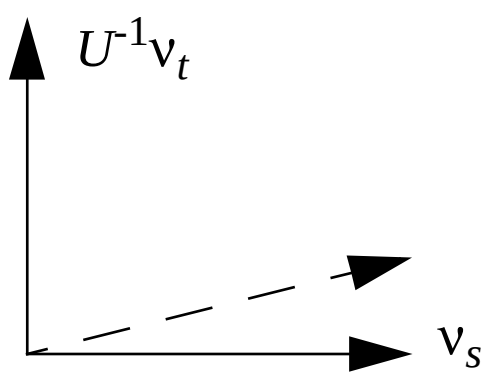

$$Q\begin{bmatrix} v_s \\ U^{-1}v_t \end{bmatrix} = \begin{bmatrix} (1 - 4|U_{ts}|^2) & 2U_{ts} \\ -2U_{ts}^* & 1 \end{bmatrix}\begin{bmatrix} v_s \\ U^{-1}v_t \end{bmatrix}$$

where $Q \equiv -I_s U^{-1} I_t U$

**Figure: The operator $Q \equiv -I_s U^{-1} I_t U$ preserves the 2 dimensional vector space spanned by $v_s$ & $U^{-1}v_t$, it rotates each vector in the space by approximately $2|U_{ts}|$ radians. Since $v_s$ & $U^{-1}v_t$ are almost orthogonal, in approximately $\frac{\pi}{4|U_{ts}|}$ applications, it transforms $v_s$ into $U^{-1}v_t$.**

## 4. Examples of quantum mechanical algorithms

The interesting feature of the analysis of section 3 is that $U$ can be *any* unitary transformation - furthermore *s* and *t* can be *any* two basis states. Clearly, it can be used to design algorithms where $U$ is a transformation in a quantum computer - this paper gives five such applications. The first three are search-type applications, the next two are statistical applications. Additional applications are presented in [Qntappl].

The following general approach is used in each of the algorithms. The problem is reduced to one of getting the system into a state *t*. A unitary transform $U$ and the initial state *s* are selected and $U_{ts}$ is calculated. By section 3, it follows that if we start with initial state *s* and carry out $O\left(\frac{1}{|U_{ts}|}\right)$ repetitions of the operation sequence $-I_s U^{-1} I_t U$ followed by a single application of $U$, there is an appreciable amplitude of the system reaching the *t*-state.

The analyses and derivation of the algorithms is considerably simpler than comparable analyses and derivation of existing algorithms. The reason for this is that the result of section 3 derives the amplitude amplification based just on $U_{ts}$ and the unitarity of the matrix $U$, i.e. it does not require us to keep track of the amplitudes in the remaining states during intermediate steps.

This power is most evident in the derivation of the $O\left(\frac{1}{\varepsilon}\right)$ step algorithm to estimate the mean to a precision $\varepsilon$ ((v) in this section). Anyone who has looked at the paper [Mean] that derived a, much more complicated, $O\left(\frac{1}{\varepsilon^{1.5}}\right)$ step algorithm for the same problem will immediately appreciate the usefulness of this approach.

### (i) Exhaustive search starting from the $\bar{0}$ state

The $N = 2^n$ states to be searched are represented by $n$ qubits. In case the starting state *s* be the $\bar{0}$ state and the unitary transformation $U$ is chosen to be $W$ (the W-H transformation as discussed in section 2), then $U_{ts}$ for any target state *t* is $\frac{1}{\sqrt{N}}$. This section gives an algorithm requiring $O\left(\frac{1}{|U_{ts}|}\right)$, i.e. $O(\sqrt{N})$ steps to reach the *t*-state with certainty. Although the derivation is different, the end result is the same quantum search algorithm as [Search1] and [Search2].

As indicated in section 3, this algorithm starts with the with $\bar{0}$ state carries out repeated operations of $Q \equiv -I_s U^{-1} I_t U$. With $U^{-1} = U = W$ and $s$ as the $\bar{0}$ state, $Q$ becomes $Q \equiv -I_{\bar{0}} W I_t W$; hence the operation sequence is: $\ldots(-I_{\bar{0}} W I_t W)(-I_{\bar{0}} W I_t W)(-I_{\bar{0}} W I_t W)\ldots$ . By rearranging parentheses and shifting minus signs, this may be seen to be alternating repetitions of $-W I_{\bar{0}} W$ and $I_t$.

The operation sequence $-W I_{\bar{0}} W$ is simply the *inversion about average* operation [Search2]. To see this, write $I_{\bar{0}}$ as $I - 2 v_{\bar{0}} v_{\bar{0}}^T$. Therefore for any vector $\bar{x}$:

$$-W I_{\bar{0}} W \bar{x} = -W(I - 2 v_{\bar{0}} v_{\bar{0}}^T) W \bar{x} = -\bar{x} + 2 W v_{\bar{0}} v_{\bar{0}}^T W \bar{x}$$

. It is easily seen that $W v_{\bar{0}} v_{\bar{0}}^T W \bar{x}$ is another vector each of whose components is the same and equal to $A$ where $A \equiv \frac{1}{N} \sum_{i=1}^{N} x_i$ (the average value of all components). Therefore the $i^{th}$ component of $-W I_{\bar{0}} W \bar{x}$ is simply: $(-x_i + 2A)$. This may be written as $A + (A - x_i)$, i.e. each component is as much above (below) the average as it was initially below (above) the average, which is precisely the inversion about average [Search2].

**(ii) Exhaustive search starting from an arbitrary basis state**

In case $U$ be chosen as the W-H transform $W$, the matrix element $U_{ts}$ between *any* pair of states *s* and *t* is $\pm\frac{1}{\sqrt{N}}$. Therefore the search algorithm can start with any of the $N = 2^n$ basis states as the initial state *s* and the procedure described in section 3, yields an algorithm to reach *t* in $O\left(\frac{1}{|U_{ts}|}\right)$, i.e. $O(\sqrt{N})$, iterations. Therefore instead of starting with the $\bar{0}$ state, as in (i), we could equally well start with *any* basis state *s*, and repeatedly apply the operation sequence $Q = -I_s W I_t W$ to obtain an equally efficient $O(\sqrt{N})$ algorithm.

The dynamics is similar to (i); only there is no longer the convenient inversion about average interpretation. The approach of section 3 makes it possible to derive this algorithm since we just need to obtain the values of $U_{ts}$ for a single application of $U$, whereas in [Search1][Search2] we had to take the analysis all the way to the end based on the inversion about average interpretation.

**(iii) Search when an item *near* the desired state is known**

***Problem*** Assume that an *n* bit word is specified - the desired word differs from this in exactly *k* bits.

***Solution*** The effect of this constraint is to reduce the size of the solution space. One way of making use of this constraint, would be to map this to another problem which exhaustively searched the reduced space using (i) or (ii). However, such a mapping would involve additional overhead. This section presents a different approach which also carries over to more complicated situations as discussed in [Qntappl].

Instead of choosing $U$ as the W-H transform, as in (i) and (ii), in this algorithm $U$ is tailored to the problem under consideration. The starting state *s* is chosen to be the specified word. The operation $U$ consists of the transformation $\begin{bmatrix} \sqrt{1-\frac{1}{\alpha}} & \frac{1}{\sqrt{\alpha}} \\ \frac{1}{\sqrt{\alpha}} & -\sqrt{1-\frac{1}{\alpha}} \end{bmatrix}$, applied to each of the *n* qubits ($\alpha$ is a variable parameter yet to be determined) - note that if $\alpha$ is $2$, we obtain the W-H transform of section 2. Calculating $U_{ts}$ it follows that $|U_{ts}| = \left(1-\frac{1}{\alpha}\right)^{\frac{n-k}{2}}\left(\frac{1}{\alpha}\right)^{\frac{k}{2}}$, this is maximized when $\alpha$ is chosen as $\frac{n}{k}$ ;then $\log|U_{ts}| = \frac{n}{2}\log\frac{n-k}{n} - \frac{k}{2}\log\frac{n-k}{k}$

The algorithm of section 3 can now be used - as in (i) and (ii), this consists of repeating the sequence of operations $-I_s U^{-1} I_t U$, $O\left(\frac{1}{|U_{ts}|}\right)$ times, followed by a single application of $U$.

The size of the space being searched in this problem is ${}^nC_k$ which is equal to $\frac{n!}{n-k!k!}$. Using Stirling's approximation: $\log n! \approx n\log n - n$, from this it follows that $\log {}^nC_k \approx n\log\frac{n}{n-k} - k\log\frac{k}{n-k}$, comparing this to the number of steps required by the algorithm, we find that the number of steps in this algorithm, as in (i) and (ii), varies approximately as the square-root of the size of the solution space being searched.

**(iv) Estimating the median to a precision $\varepsilon$**

Assume that we are given $N = 2^n$ values between 0 and 1 denoted by $x_0, x_1 \ldots x_{N-1}$ and a certain threshold $\theta$. Let $\varepsilon$ denote the fractional difference in the number of values above and below $\theta$, i.e. the number of values below $\theta$ is $\frac{N}{2}(1-\varepsilon)$. Given the bound $|\varepsilon| < 2\varepsilon_0$, the task is to find an estimate $\varepsilon_e$ such that $\left||\varepsilon_e| - |\varepsilon|\right| < \frac{\varepsilon_0}{4}$.

If we can solve the above problem in $O\left(\frac{1}{\varepsilon_0}\right)$ steps, then it is possible to iterate with different thresholds and estimate the median to a precision $\varepsilon$ in $O\left(\frac{1}{\varepsilon}\right)$ steps [Median]. Classically it needs $\Omega\left(\frac{1}{\varepsilon^2}\right)$ steps to estimate the median with a precision of $\varepsilon$. (All bounds in this and the next example are upto polylogarithmic factors).

**Solution:** This sub-section gives a scheme for the estimation of $\varepsilon$ with an error bound of $\frac{\varepsilon_0}{4}$ in $O\left(\frac{1}{\varepsilon_0}\right)$ steps. The basic idea is to devise a unitary operation $U$ due to which the amplitude in a *t*-state comes out to be proportional to the statistic to be estimated. By repeating the operation $-I_s U^{-1} I_t U$, an appropriate number of times, the probability in the *t*-state is boosted to the point that it can be estimated by carrying out the experiment a few times.

Consider an $N = 2^n$ state quantum system represented by $n$ qubits - associate a value $x_\alpha$ with each state, $S_\alpha$. Next consider the unitary transform $R$ which is a selective inversion operation (as discussed in section 2):

$R$: *In case the value $x_\alpha$ associated with the state $S_\alpha$ is smaller than the threshold $\theta$, invert the amplitude in $S_\alpha$.*

Start the system in the $\bar{0}$ state and consider the unitary operation $U = WRW$. It is easily seen that after $U$ the amplitude of the system being in the $\bar{0}$ state is $\varepsilon$. Also note that $U^{-1} = W^{-1}R^{-1}W^{-1} = WRW$. Therefore, by the analysis of section 3, it follows that after $O\left(\frac{1}{\varepsilon_0}\right)$ repetitions of $Q \equiv -I_{\bar{0}} U^{-1} I_{\bar{0}} U$, followed by a single application of $U$, the amplitude in the $\bar{0}$ state reaches $\frac{\varepsilon}{4\varepsilon_0}$. Now if a measurement be made which projects the system onto one of its basis states, the probability of getting $\bar{0}$ is $\left(\frac{\varepsilon}{4\varepsilon_0}\right)^2$. It follows from the central limit theorem [Feller] that by repeating this experiment $O(M)$ times, it is possible to estimate $\left|\frac{\varepsilon}{\varepsilon_0}\right|$ with a precision of $\frac{1}{\sqrt{M}}$, and hence $|\varepsilon|$ with a precision $\frac{\varepsilon_0}{\sqrt{M}}$. By appropriately choosing $M$, it is possible to estimate $|\varepsilon|$ within an error bound of $\frac{\varepsilon_0}{4}$.

**(v) Estimating the mean to a precision $\varepsilon$**

Assume that we are given $N = 2^n$ values denoted by $x_0, x_1 \ldots x_{N-1}$, each $x_\alpha$ lies in the range $(-0.5, 0.5)$ - the task is to estimate the mean (denoted by $\mu$) to a specified precision $\varepsilon$. Classically this would take $\Omega\left(\frac{1}{\varepsilon^2}\right)$ steps. The best known quantum mechanical algorithm takes $O\left(\frac{1}{\varepsilon^{1.5}}\right)$ steps [Mean]. This section presents a simple $O\left(\frac{1}{\varepsilon}\right)$ step quantum mechanical algorithm (upto polylogarithmic factors).

Start with a relatively large $\varepsilon_0$ which is chosen so that $|\mu| < \varepsilon_0$. Carry out the following loop (i)...(iii):

(i) Estimate $\mu$ with a precision of $\varepsilon_0$, i.e. find an estimate $\mu_e$ such that $|\mu_e - \mu| < \frac{\varepsilon_0}{2}$. This step is described in the "Main algorithm" below.

(ii) Shift each of the numbers by the newly estimated mean, i.e. $x_\alpha = x_\alpha - \mu_e$.

(iii) If $\varepsilon_0 > \varepsilon$, replace $\varepsilon_0$ by $\frac{\varepsilon_0}{2}$ and go to (i).

The mean may be estimated as the sum of the estimated $\mu_e$ in each iteration of step (i) of the loop.

**Main algorithm** The heart of the above algorithm is step (i) of the above loop, i.e. estimating $\mu$ with a precision of $\varepsilon_0$ when $|\mu| < \varepsilon_0$. Consider a $(2^{n+1} + 1)$ state quantum mechanical system represented by $(n + 2)$ qubits with the following encoding for the states:

$S_0 \quad S_1 \quad \ldots\ldots\ldots \quad S_{N\text{-}1}$

(first two bits are 00, next *n* bits indicate the $S_\alpha$ state)

$R_0 \quad R_1 \quad \ldots\ldots\ldots \quad R_{N\text{-}1}$

(first two bits are 01, next *n* bits indicate the $R_\alpha$ state)

Q

(first bit is a 1, the next $(n + 1)$ bits are zero).

The $S_\alpha$ states are the computational states, for each $S_\alpha$ state there is an $R_\alpha$ state, also there is a single Q state. Each state is encoded by (n+2) qubits as shown above.

Associate the value $x_\alpha$ with each of the $R_\alpha$ and $S_\alpha$ states. We need the following 4 unitary operations in this algorithm (these are denoted by $M_1, M_2, R_1$ and $W_1$.) It may be verified that these are indeed valid unitary operations and, with the above encoding, they require operations on only a few qubits at a time.

$M_1$

*If in state* $S_0$: go to Q with an amplitude of $\frac{1}{2}$, stay in $S_0$ with an amplitude of $\frac{\sqrt{3}}{2}$;

*if in state* Q: go to $S_0$ with an amplitude of $\frac{1}{2}$, stay in Q with an amplitude of $-\frac{\sqrt{3}}{2}$;

*if in any other state*: stay in the same state.

$M_2$

*If in state* $S_0$: go to Q with an amplitude of $\frac{1}{\sqrt{2}}$, stay in $S_0$ with an amplitude of $\frac{1}{\sqrt{2}}$;

*if in state* Q: go to $S_0$ with an amplitude of $-\frac{1}{\sqrt{2}}$, stay in Q with an amplitude of $\frac{1}{\sqrt{2}}$;

*if in any other state:* stay in the same state.

$R_1$

*If in state* $S_\alpha$: go to $R_\alpha$ with an amplitude of $\sqrt{\frac{2}{3} - \frac{4}{3}x_\alpha^{\ 2}}$, stay in $S_\alpha$ with an amplitude of $\left(\frac{1}{\sqrt{3}} + \frac{2ix_\alpha}{\sqrt{3}}\right)$;

*if in state* $R_\alpha$: go to $S_\alpha$ with an amplitude of $\sqrt{\frac{2}{3} - \frac{4}{3}x_\alpha^{\ 2}}$, stay in $R_\alpha$ with an amplitude $\left(-\frac{1}{\sqrt{3}} + \frac{2ix_\alpha}{\sqrt{3}}\right)$;

*if in any other state*: stay in the same state.

| | | |
|---|---|---|
| $W_1$ | *If in state* $S_\alpha$: | carry out the W-H transform on the $S_\alpha$ states; |
| | *if in any other state*: | stay in the same state. |

It is easily shown that if we start with the $S_0$ state and carry out the operation $U$, where $U$ is defined as: $U \equiv M_1 W_1 R_1 W_1 M_2$, then the amplitude in the $S_0$ state will be $\frac{i\mu}{\sqrt{2}}$. As in the median estimation, (iv), $U^{-1}$ is the product of the inverses of the same matrices as $U$, but in the opposite order. By the analysis of section 3, it follows that: By $O\left(\frac{1}{\varepsilon_0}\right)$ repetitions of $Q$ where $Q \equiv -I_{S_0} U^{-1} I_{S_0} U$ followed by a single application of $U$, the amplitude in the $S_0$ state becomes $\frac{i\mu}{\sqrt{2}\varepsilon_0}$. Now if a measurement is made which projects the system onto one of its basis states, the probability of getting the $S_0$ state is $\frac{1}{2}\left(\frac{\mu}{\varepsilon_0}\right)^2$. It can be shown by the central limit theorem that by repeating this entire experiment $O(M)$ times, it is possible to estimate $\frac{\mu}{\varepsilon_0}$ with a precision of $\frac{1}{\sqrt{M}}$. By an appropriate choice of $M$, it is possible to estimate $\mu$ within an error bound of $\frac{\varepsilon_0}{2}$.

Note that in the above analysis we only needed to design the unitary transform $U$ to appropriately adjust the one matrix element that we cared about - just the condition that it was unitary took care of everything else.

**5. General quantum mechanical algorithms**

The framework described in this paper can be used to enhance the results of *any* quantum mechanical algorithm. Assume there is a quantum mechanical algorithm $Q$ due to which there is a finite amplitude $Q_{ts}$ for transitions from the starting state *s* to the target state *t*. The probability of being in the state *t* is hence $|Q_{ts}|^2$ - it will therefore take $\Omega\left(\frac{1}{|Q_{ts}|^2}\right)$ repetitions of $Q$ to get a single observation of state *t*. Since the quantum mechanical algorithm $Q$ is a sequence of $\eta$ elementary unitary operations: $Q_1 Q_2 \ldots Q_\eta$, it is itself a unitary transformation. Also, $Q^{-1} = Q_\eta^{-1} Q_{\eta-1}^{-1} \ldots Q_1^{-1}$. The inverse of any elementary unitary operation on a small number of qubits is another elementary unitary operation on the same qubits and can hence be synthesized. Applying the framework of section 3, it follows that by starting with the system in the *s* state and repeating the sequence of operations: $-I_s Q^{-1} I_t Q$, $O\left(\frac{1}{|Q_{ts}|}\right)$ times followed by a single application of $Q$, it is possible to reach the *t*-state with certainty.

**6. Conclusion**

Designing a useful quantum computer has been a daunting task for two reasons. First, because the physics to implement this is different from what most known devices use and so it is not clear what its structure should be like, The second reason is that once such a computer is built, few applications for this are known where it will have a clear advantage over existing computers. This paper has given a general framework for the synthesis of a category of algorithms where the quantum computer would have an advantage. The main concept is that if there is a finite amplitude for a certain transition, this can amplified by a factor of $M$ in $M$ iterations of the algorithm. Several applications were presented. It is expected that this formalism will also be useful in the physical design of quantum computers, since it demonstrates that quantum mechanical algorithms can be implemented through general properties of unitary transformations and can thus adapt to available technology.

## 7. Acknowledgments

Thanks to Charles Bennett, Andy Yao, Eddie Farhi, Sam Gutmann and particularly Norm Margolus for their quick and constructive feedback.